\documentclass[aps,prl,twocolumn,10pt,superscriptaddress]{revtex4-1}
\usepackage{amsmath}
\usepackage{amssymb}
\usepackage{amsfonts}
\usepackage{color}
\usepackage{graphics}
\usepackage[pdftex]{graphicx}
\usepackage[utf8x]{inputenc}
\usepackage[colorlinks=true]{hyperref}
\usepackage{footmisc}
\usepackage{braket}

\newcommand{\paren}[1]{{\left({#1}\right)}}

\newcommand{\Na}{\mathrm{Na}}
\newcommand{\Cs}{\mathrm{Cs}}

\newcommand{\todo}[1]{}

\ifpdf
\pdftrailerid{}
\hypersetup{
  pdfcreator={},
  pdfproducer={}
}
\fi

\newcommand{\harvardphysics}{\affiliation{Department of Physics, Harvard University, Cambridge, Massachusetts 02138, USA}}
\newcommand{\harvardccb}{\affiliation{Department of Chemistry and Chemical Biology, Harvard University, Cambridge, Massachusetts 02138, USA}}
\newcommand{\cua}{\affiliation{Harvard-MIT Center for Ultracold Atoms, Cambridge, Massachusetts 02138, USA}}
\newcommand{\agendile}{\affiliation{Agendile LLC, Cambridge, Massachusetts 02139, USA}}
\newcommand{\itamp}{\affiliation{ITAMP, Harvard-Smithsonian Center for Astrophysics, Cambridge, Massachusetts 02138, USA}}
\newcommand{\gradstudent}{
  \harvardphysics
  \harvardccb
  \cua
}

\usepackage{xcolor}
\newcounter{TRC}

\begin{document}
\title{Coherent optical creation of a single molecule}
\author{Yichao~Yu}
\thanks{Y.Y.~and K.W.~contributed equally to this work.}
\gradstudent
\author{Kenneth~Wang}
\thanks{Y.Y.~and K.W.~contributed equally to this work.}
\gradstudent
\author{Jonathan~D.~Hood}
\affiliation{Department of Chemistry, Purdue University, West Lafayette, Indianna, 47906, USA}
\author{Lewis~R.~B.~Picard}
\gradstudent
\author{Jessie~T.~Zhang}
\gradstudent
\author{William~B.~Cairncross}
\harvardccb
\harvardphysics
\cua
\author{Jeremy~M.~Hutson}
\affiliation{Joint Quantum Centre Durham-Newcastle, Department of Chemistry, Durham University, Durham, DH1 3LE, United Kingdom}
\author{Rosario Gonzalez-Ferez}
\affiliation{Instituto Carlos I de F\'{\i}sica Te\'orica y Computacional, and Departamento de F\'{\i}sica At\'omica, Molecular y Nuclear,  Universidad de Granada, 18071 Granada, Spain}
\itamp
\author{Till Rosenband}
\agendile
\author{Kang-Kuen~Ni}
\email[To whom correspondence should be addressed: ]{ni@chemistry.harvard.edu}
\harvardccb
\harvardphysics
\cua

\date{\today}

\begin{abstract}
  We report coherent association
  of atoms into a single weakly bound NaCs molecule in an optical tweezer
  through an optical Raman transition.
  The Raman technique uses a deeply bound electronic excited intermediate state
  to achieve a large transition dipole moment while reducing photon scattering.
  Starting from two atoms in their relative motional ground state,
  we achieve an optical transfer efficiency of $69\mathrm{\%}$.
  The molecules have a binding energy of $770.2~\mathrm{MHz}$ at $8.83(2)~\mathrm{G}$.
  This technique does not rely on Feshbach resonances or narrow excited-state lines
  and may allow a wide range of molecular species to be assembled atom-by-atom.
\end{abstract}

\maketitle

Diverse species of fully quantum-controlled ultracold molecules are desired
for a variety of applications including precision measurements~\cite{
  Kondov2019,Nick_and_Ivan2017, PhysRevA.101.042504, Andreev2018,
  PhysRevLett.119.153001, hudson2011},
quantum simulations~\cite{Micheli2006, Yao2018, Wall2015, wall2015realizing},
quantum information processing~\cite{DeMille2002, Ni2018, Hudson2018, Lin2019},
and studies of ultracold chemistry~\cite{Bala2016,Hu1111,Segev2019,deJongh626}.
While many innovative approaches in the last few years
have directly cooled different species of
molecules
below 1~mK~\cite{Norrgard2016,Prehn2016,Truppe2017,Anderegg2018,  PhysRevX.10.021049,
  Mitra1366},
the highest phase-space-density gas~\cite{Demarco2018} and
trapped individual molecules~\cite{Zhang2020,He331}
have been achieved through the association of ultracold atoms.

Molecular association of ultracold atoms takes advantage of the cooling and trapping techniques
that have been developed for atoms.
Associating atoms into deeply bound molecules is challenging
because of the small wavefunction overlap between the free-atom and molecular states
and the release of large binding energy.
A widely used method of overcoming these challenges is to associate atom pairs
into weakly bound molecules first,
and then transfer the molecules from this single internal state
to a desired rovibrational and electronic state,
releasing the binding energy by stimulated emission~\cite{Danzl2008, Ni2008, Lang2008,
  Takekoshi2014, Molony2014, Park2015, Guo2016, Kondov2019, Voges2020}.
So far, molecular association has generally been achieved by magnetoassociation
through a magnetic Feshbach scattering resonance.
Exceptions include Sr$_2$, where narrow-linewidth ($\sim 20~\mathrm{kHz}$) excited states
are available and optical association can be driven coherently~\cite{Stellmer2012,Reinaudi2012},
and $^{87}$Rb$^{85}$Rb with molecular states bound by only $1-2~\mathrm{MHz}$~\cite{He331}.
With these requirements, molecules involving non-magnetic atoms~\cite{PhysRevX.10.031037}
or atoms without narrow intercombination lines remain difficult to associate.

\begin{figure*}
  \includegraphics[width=\textwidth]{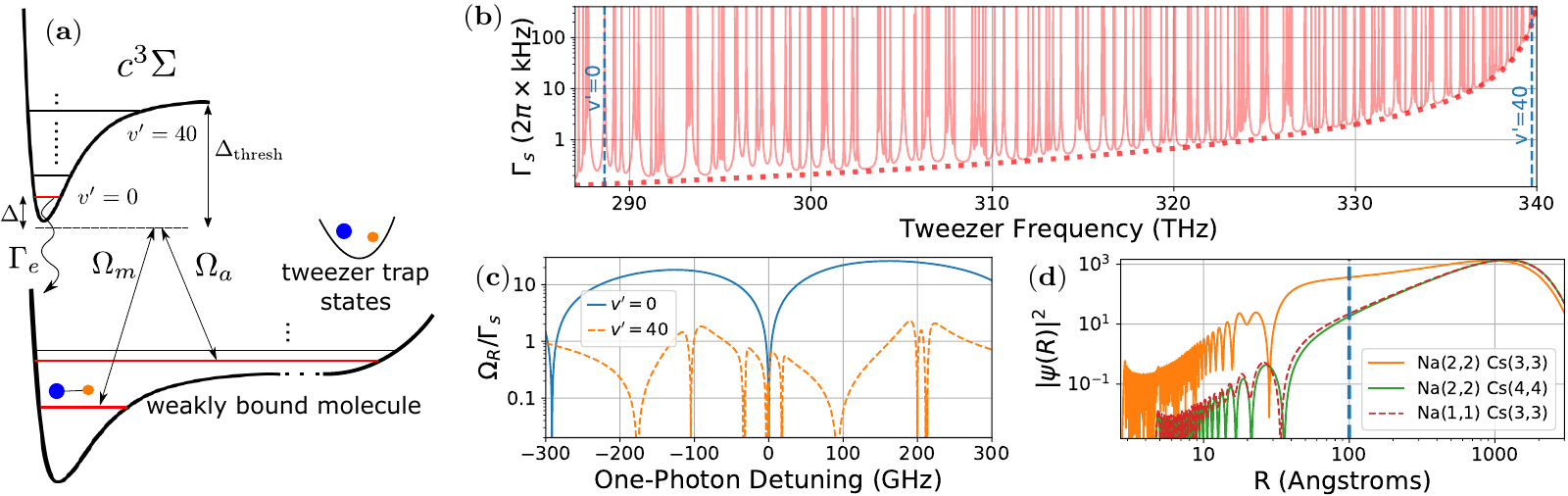}
  \caption{Optical creation of single molecules from single atoms in an optical tweezer.
    (a) Schematic of the optical transition from an atom pair to a weakly bound molecule.
    The initial state is the relative motional ground state between the two atoms
    and the final state is the first molecular bound state.
    The transition is driven by a pair of laser frequencies whose difference ($< 1$~GHz) matches the molecular binding energy.
    The lasers are detuned by $\Delta$ from an excited vibrational sublevel
    in the $\mathrm{c^3\Sigma^+(\Omega = 1)}$ electronic state
    in order to reduce scattering during the transfer.
    (b) Calculation of the scattering rate vs. laser frequency
    including all vibrational states of 8 excited-state potentials
    and the atomic continuum.
    The assumed excited-state linewidth for molecular lines is $50~\text{MHz}$.
    The dotted red line shows the contribution from the near-threshold states, which approximately scales as $1/\Delta_{\mathrm{thresh}}^2$.
    (c) Comparison between vibrational sublevels of the intermediate excited state
    for the Raman transition.
    The lowest sublevel (blue)
    has a larger optimum ratio of Raman Rabi frequency to scattering rate
    than the higher sublevel (orange).
    (d) The large scattering length for $\Na(2,2),\Cs(3,3)$ is associated with enhancement of the relative wavefunction at short
    internuclear distance ($R$).
    \label{f-theory}
  }
\end{figure*}

Here, we demonstrate coherent association of an atom pair to a weakly bound molecule
by two-photon optical Raman transfer via an electronic excited state,
as shown in Fig.~\ref{f-theory}a, taking NaCs as the prototype system.
The technique does not rely on a Feshbach resonance,
molecular states bound by a only few~MHz, or a narrow excited state.
The resulting single molecule is in a well-defined internal quantum state
and predominantly in its motional ground state.
A vibrational state of the electronic excited state $\mathrm{c^3\Sigma^+}(\Omega = 1)$
serves as the intermediate state in the Raman technique,
and is chosen to minimize photon scattering during Raman Rabi oscillations.
To reduce photon scattering and sensitivity to laser intensity noise further,
we choose the initial and final states that balance the two Rabi frequencies as much as possible.
This approach applies to a variety of molecules that can be created atom-by-atom
with full quantum-state control.

The optical Raman transfer is illustrated
by the idealized three-level system shown in Fig.~\ref{f-theory}a,
where the initial atomic state and the target weakly bound molecular state are coupled to an intermediate state by two lasers with Rabi frequencies $\Omega_{\rm a}$ and $\Omega_{\rm m}$, one-photon detuning $ \Delta $, and all Rabi frequencies are population oscillation frequencies.
The transfer Raman Rabi frequency is given by $\Omega_{\rm a}\Omega_{\rm m} / (2\Delta)$~\cite{Wineland2003}.
Unlike Raman transitions in atoms, the one-photon Rabi frequencies are greatly imbalanced ($\Omega_{\rm a}/\Omega_{\rm m} \ll 1$)
due to the small wavefunction overlap between the atomic state and the intermediate state,
and scattering losses are dominated by the final state. Because the energy difference between the atomic state and target molecular state is small ($ < 1~\mathrm{GHz} $) compared to the single-photon detuning of $80$ to $200$~GHz, both beams scatter nearly equally with a total rate $ \Gamma_{\rm e} \Omega_{\rm m}^2 / (2\Delta^2)$, where $ \Gamma_{\rm e} $ is the excited-state linewidth.  The two beams have equal power to maximize the Raman Rabi frequency at fixed total power.
In this idealized treatment, the ratio between the Raman Rabi frequency and the scattering rate is $ \Omega_{\rm a}/\Omega_{\rm m} \times \Delta/\Gamma_{\rm e} $, which limits the transfer efficiency into the molecular state. At the same time, the intensity-stability requirement is determined by the ratio of Raman Rabi frequency to light shift $\Omega_{\rm a}/\Omega_{\rm m}$. Notably, both figures of merit improve with a larger ratio $\Omega_{\rm a}/\Omega_{\rm m}$.

Earlier experiments used excited states with high vibrational quantum number for the intermediate state
of the Raman transition to ensure a large Raman Rabi frequency~\cite{Wynar2000,Rom2004}.
However, a complete picture includes both the many vibrational levels
of the excited electronic state and the atomic continuum.
Then the scattering and Raman Rabi rates are sums over all possible intermediate states.
As there is large overlap between the target molecular state and other vibrationally excited states, intermediate states that are closer to the dissociation threshold result in a large scattering rate.
This scattering is approximately proportional to $1/\Delta_{\mathrm{thresh}}^2$,
where $\Delta_{\mathrm{thresh}}$ is the detuning from the dissociation threshold,
which is smaller for intermediate states that are deeply bound.

We optimize over intermediate states by calculating the total Raman Rabi frequency $\Omega_{\rm R}$
and scattering rate $\Gamma_{\rm s}$ at different detunings from the atomic threshold,
taking into account all states of
8 excited molecular electronic potentials~\cite{Korek2007, Grochola2011, Zaharova2009, Grochola2010, Zabawa2012}
and the continuum~\cite{Liu2017};
see Supplementary Material for details.
This calculation shows that the figure of merit $\Omega_{\rm R}/\Gamma_{\rm s}$
can be larger for the lowest vibrational state compared to higher bound states
at a cost of a smaller transfer rate $\Omega_{\rm R}$, as shown in Fig.~\ref{f-theory}c.
As a result, we choose the $v'=0$ level of $\mathrm{c^3\Sigma^+}(\Omega = 1)$
as the intermediate state near which to drive Raman transitions.

In addition to the intermediate state,
the choice of initial and final Zeeman and hyperfine states affects the single-photon rates $\Omega_{\rm a}$ and $\Omega_{\rm m}$.
Due to the small extent of the intermediate-state wavefunction
compared to that of the trapped atoms,
$\Omega_{\rm a}$ is approximately proportional to
the amplitude of the relative atomic wavefunction at short distance,
within the range of the molecular potential.
To increase this amplitude, one can increase the external confinement of atom pairs.
In a harmonic approximation
the short-range amplitude is proportional to $ \omega_{\text{trap}}^{3/4} $ or $P^{3/8}$,
where $ \omega_{\text{trap}} $ is the trap frequency and $P$ is the optical power in the tweezer trap \cite{Mies2000} with a fixed beam waist. However, additional power may not be available and also leads to additional undesired scattering.
Alternatively, one can choose an atomic pair state with a large scattering length
(positive or negative).
For such states, the amplitude of the relative atomic wavefunction is substantially enhanced
at short range, as shown in Fig.~\ref{f-theory}d.
For our system of Na and Cs atoms,
we choose a spin-state combination $\ket{\uparrow_{\Na} \downarrow_{\Cs}}\equiv \ket{f=2,m_f=2}_{\Na}\ket{f=3,m_f=3}_{\Cs}$ that has a large and negative scattering length of
$a(\uparrow_{\Na} \downarrow_{\Cs}) \approx -700a_0$~\cite{Hood2019}.
All other stable spin combinations give smaller scattering lengths ($<~50~a_0$).

To identify a suitable target molecular state, we carry out coupled-channel calculations of the near-threshold bound states, as described in the Supplemental Material.
Choosing a bound state with similar spin character to the atomic state  minimizes the sensitivity of the transition frequency to magnetic field.
A suitable state with this character is predicted about 763 MHz below the $\ket{\uparrow_{\Na} \downarrow_{\Cs}}$ threshold and the ratio $\Omega_{\rm a}/\Omega_{\rm m}$ increases to about 0.013.
Compared to $\Omega_{\rm a}/\Omega_{\rm m} \approx 0.003$
for other combinations, this relaxes the intensity stability requirement to the percent level and enhances the Raman Rabi frequency.

\begin{figure}
  \includegraphics[width=0.5\textwidth]{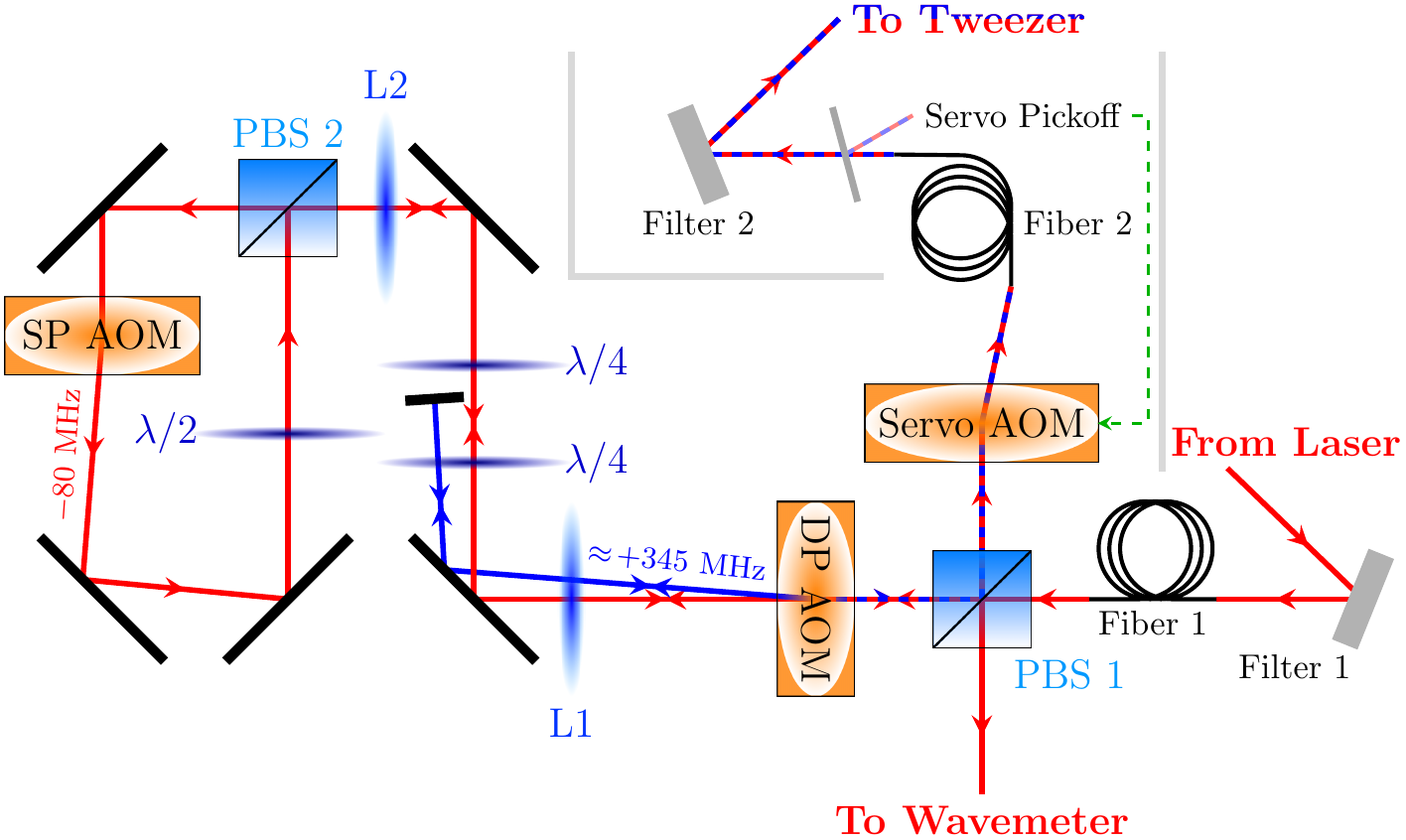}
  \caption{
    Beam path for generating two frequencies in the tweezer for the Raman transition (only essential optics shown.)
    Two filters, one directly after the laser source and one after both optical fibers clean the light from a fiber amplifier seeded by an external cavity diode laser.
    The red beam path is the always-present $0^{\mathrm{th}}$ order of the double pass~(DP) AOM.
    The blue beam path is the switchable $1^{\mathrm{st}}$ order of the DP AOM. It generates the second frequency for the Raman transition.
    To reduce interferences that cause relative power fluctuation of the two frequencies, the $0^{\mathrm{th}}$ order is frequency shifted by $-80$~MHz before recombination.
    The experiment typically starts with the single-pass (SP) AOM on and the DP AOM off.
    When driving the Raman transition, the RF powers for both AOMs are ramped simultaneously
    to achieve the desired optical power at both frequencies, while keeping the total optical power fixed.
    \label{f-beampath}
  }
\end{figure}

Experimentally, we first prepare two atoms in a well-defined external and internal quantum state
by using techniques developed previously~\cite{Liu2018, Liu2019, Wang2019}.
In brief, the experimental cycle begins by stochastically loading a single ${}^{23}\Na$ atom
and a single ${}^{133}\Cs$ atom into separate optical tweezers.
The atoms are initially imaged to post-select loading of both atoms vs.
none or one atom.
After imaging, we turn on a $8.83(2)~\mathrm{G}$ magnetic field to define the quantization axis
for the state preparation and molecule formation steps.
Raman sideband cooling then prepares both atoms simultaneously
in the 3-dimensional motional ground state of their optical tweezers, leaving the atoms in the spin state~$\ket{\uparrow_{\Na}\uparrow_{\Cs}}\equiv \ket{f=2,m_f=2}_{\Na}\ket{f=4,m_f=4}_{\Cs}$,
which has a small scattering length.
The weak two-atom interaction allows merging of the two tweezers with minimum pertubation so that they remain in the motional ground state.

\begin{figure}[t!]
  \includegraphics[width=0.48\textwidth]{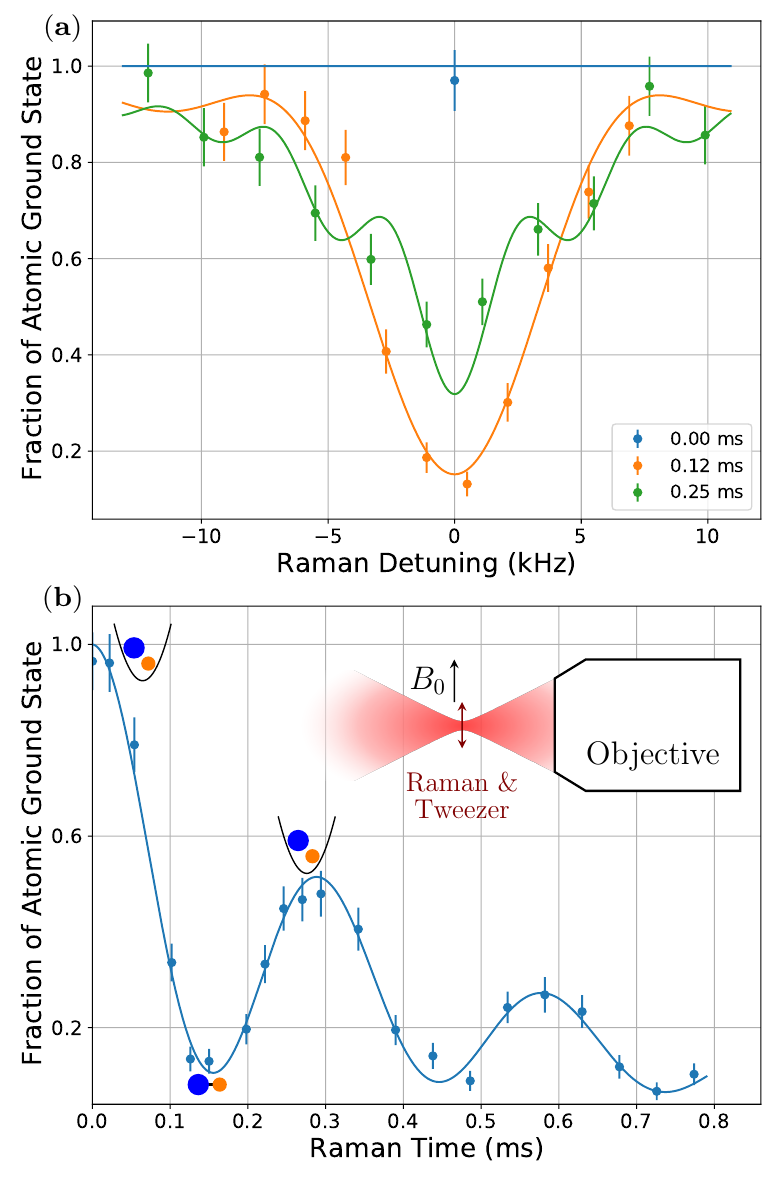}
  \caption{Coherent transfer of atoms to molecules.  The molecular state is dark to the imaging step and corresponds to zero signal.
    (a) Raman difference frequency scans for various durations
    showing the resonance as a function of the detuning from
    the fitted resonance at $770.5715(1)$~MHz, near the calculated value of $763$~MHz.
    (b) Raman pulse-length scan on resonance.
    A decaying Rabi oscillation shows the coherence of
    the Raman transfer process.
    A model is fitted to (a) and (b) to determine
    the Raman Rabi frequency and loss rates.
    \emph{Inset:} Geometry and polarization of trap and Raman beam relative to the magnetic field.
    The 3.25~mW beam is focused to a waist of $0.9~\mathrm{\mu m}$
    that confines the atoms and molecule.
    A perpendicular $B_0=8.83(2)~\mathrm{G}$ magnetic field
    defines the quantization axis and the atoms experience predominantly $\pi$-polarized light.
    \label{f-raman}}
\end{figure}

\begin{figure*}[htb]
  \includegraphics[width=\textwidth]{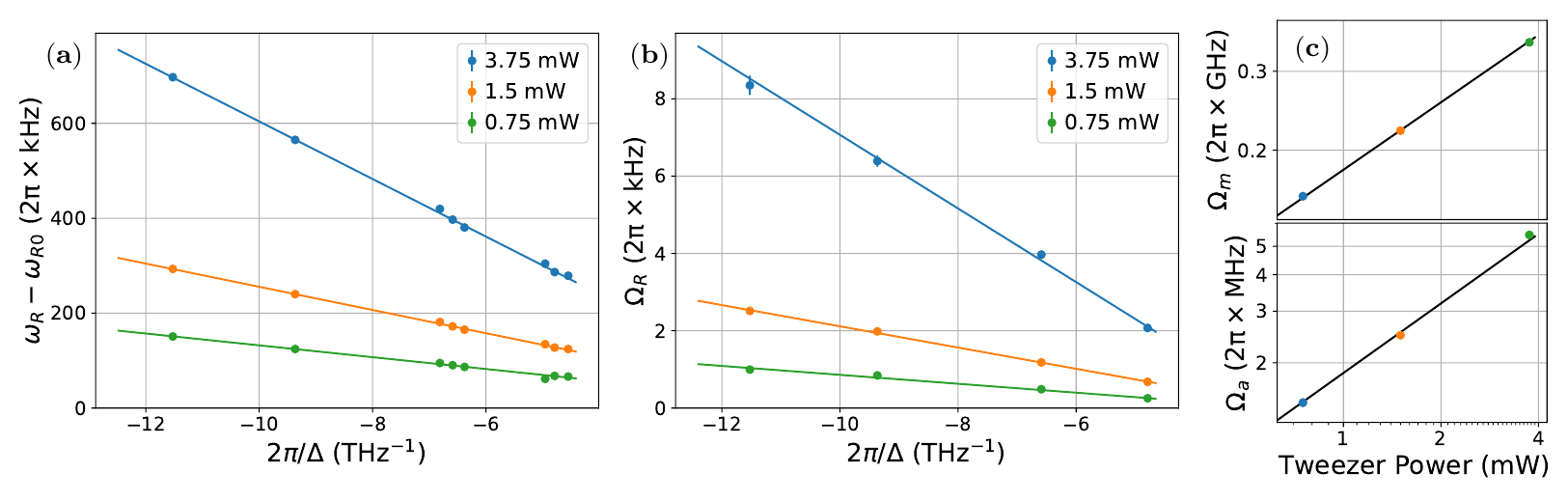}
  \caption{Raman transition parameters as a function of tweezer power and detuning.
    (a) The Raman resonance $\omega_{\rm R}$ fitted to $a_P+b_P/\Delta$, where
    $a_P$ and $b_P=(\Omega_{\rm a}^2-\Omega_{\rm m}^2)/2$
    are the power ($P$) dependent background and $v'=0$ contributions
    to the light shift and
    $\Delta\equiv2\pi\times\paren{f_{\mathrm{PA}0} - f_{\mathrm{tweezer}}}$ is the detuning from
    the $v'=0$ resonance frequency $f_{\mathrm{PA}0}$.
    $a_P$ is fitted to a model including linear and small quadratic light shift
    \todo{which assumes $\Omega_{\rm m}\gg\Omega_{\rm a}$} to obtain the Raman resonance frequency
    at zero tweezer power $\omega_{\rm R0}=2\pi\times770.1969(2)~\mathrm{MHz}$ where the statistical uncertainty is shown.
    (b) Raman Rabi frequency $\Omega_{\rm R}$ fitted to $c_P+d_P/\Delta$, where
    $c_P$ and $d_P=\Omega_{\rm a}\Omega_{\rm m}/2$
    are the background and $v'=0$ contributions that scale as $P^{1.29}$.
    The detuning is calculated from $f_{\mathrm{PA}0}$ fitted in (a). Fit parameters are listed in Table~\ref{tab:f-det:fit}.
    (c) Tweezer power dependency of $\Omega_{\rm m}$ (top) and $\Omega_{\rm a}$ (bottom) calculated from
    $b_P$ and $d_P$ on a log-log scale showing approximate $P^{0.5}$ scaling of $\Omega_{\rm m}$ and
    $P^{0.79}$ scaling of $\Omega_{\rm a}$.
    \label{f-det}}
\end{figure*}

After merging the tweezers, we drive the atoms into spin combination $\ket{\uparrow_{\Na} \downarrow_{\Cs}}$ with a large scattering length
by performing a Cs spin flip while taking into account
the $-30.7~\mathrm{kHz}$ interaction shift~\cite{Hood2019}.
This is the initial atomic state for Raman transfer.
The spin flip selectively transfers atoms in the relative motional ground state,
removing any background from atoms in excited states of relative motion
\footnote{This interaction shift is larger than the differential axial trapping frequency
  between Na and Cs atoms, which decouples the relative and center of mass motional state
  and improves the robustness of our preparation of the relative motional ground state.}.
For the experiment reported here,
$31\mathrm{\%}$ of initial two-atom population is transferred. Of this population, over $60\%$ is in the ground state of center-of-mass motion, inferred from Raman sideband thermometry.

To transfer the atom pair into the target weakly bound molecular state,
we modulate the tweezer beam with a second frequency near $770$~MHz, as shown in Fig.~\ref{f-beampath}.
The dual use of the tweezer beam for confinement and Raman transfer not only minimizes photon scattering,
but also allows a tight focus to minimize the transfer duration. A tweezer frequency far detuned (by $-151~\mathrm{GHz}$) from $v' = 0 $ ($<50$~MHz natural linewidth) reduces resonant scattering~\cite{Liu2019}.
Furthermore, two filters, each with a linewidth (FWHM) of $50~\mathrm{GHz}$,
clean the laser spectrum and prevent broadband noise from causing unwanted excitation.
As shown in Fig.~\ref{f-beampath}, one filter immediately follows the laser, while the second filter precedes the focusing objective for final cleanup of the laser spectrum.

Figure~\ref{f-raman} shows a Fourier-limited resonance together with Rabi oscillations between the atomic and molecular states.
A decaying Rabi oscillation with frequency $2\pi\times3.28(4)~\mathrm{kHz}$ fitted to the data suggests that
$69~\mathrm{\%}$ of atoms are transferred into the molecular state after a $\pi$-pulse, with the majority of molecules in the ground state of motion~\cite{Zhang2020,He331}.

\begin{figure}[ht]
  \includegraphics[width=0.48\textwidth]{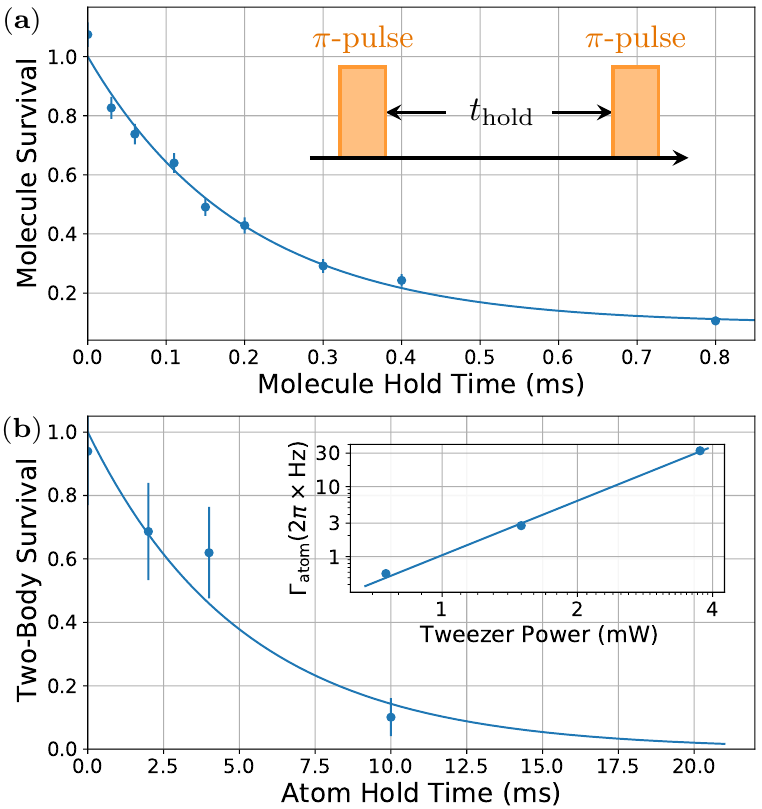}
  \caption{Lifetime measurements with about $3.25~\mathrm{mW}$ optical power.
    (a) Direct measurement of molecule lifetime.
    Molecule survival is detected by dissociating back to atoms via a second Raman transition.
    The lifetime is consistent with the decay of the Rabi oscillation in Fig.~\ref{f-raman}b.
    Inset: pulse sequence for the lifetime measurement.
    (b)~Two-body atom lifetime of $5(1)~\mathrm{ms}$
    limited by off-resonance photoassociation.
    This is used to improve the fitting of the Raman transfer data.
    Inset: Atomic scattering rate scales as
    $P_{\mathrm{tweezer}}^{2.58}\times\!2\pi\!\times1.05(6)~\mathrm{Hz/mW^{2.58}}$ on a log-log scale;
    this is consistent with a two-photon scattering process.
    \label{f-lifetime}}
\end{figure}

To understand the details and limitations of the Raman transfer process better,
we measured the properties of the two-photon resonance as a function of tweezer power and single-photon detuning.
Known dependencies of the light shift and Raman Rabi frequency $\Omega_{\rm R}$ on detuning $\Delta$ allow experimental determination of the Rabi frequencies
$ \Omega_{\rm a} $ and $\Omega_{\rm m}$ whose ratio critically affects the transfer efficiency.
Both the light shift and $\Omega_{\rm R}$ follow a $1/\Delta$ slope as shown in Fig.~\ref{f-det}a, b and include a constant offset that we attribute to coupling to other excited states that are further away in energy.
The $1/\Delta$ components due to the nearby $v'=0$ intermediate state determine $\Omega_{\rm m} $ and $ \Omega_{\rm a} $ in Table~\ref{tab:f-det:fit}.

\begin{table}[ht]
  \centering
  \begin{tabular}{|c|c|c|c|}
    $P~(\mathrm{mW})$&$0.75$&$1.5$&$3.75$\\\hline
    $f_{\mathrm{PA}0}~(\mathrm{GHz})$&\multicolumn{3}{|c|}{$288711.8$}\\\hline
    $a~(\mathrm{2\pi\times MHz})$&$770.20452(6)$&$770.2081(1)$&$770.1943(3)$\\
    $b~(\mathrm{4\pi^2\times MHz\cdot GHz})$&$-12.46(2)$&$-24.44(3)$&$-60.66(8)$\\\hline
    $c~(\mathrm{2\pi\times kHz})$&$0.29(2)$&$0.63(4)$&$2.4(2)$\\
    $d~(\mathrm{4\pi^2\times MHz\cdot GHz})$&$0.115(4)$&$0.275(6)$&$0.95(3)$\\ \hline
    $\Omega_{\rm R}~(\mathrm{2\pi\times kHz})$ & 0.49(2) & 1.18(3) & 3.97(9) \\
    $\Omega_{\rm m}~(\mathrm{2\pi\times MHz})$ & & & 348.3(3) \\
    $\Omega_{\rm a}~(\mathrm{2\pi\times kHz})$ & & & 5.5(2)
  \end{tabular}
  \caption{Fitting results for Fig.~\ref{f-det}(a,b). $\Omega_{\rm R}$ is reported at $-151~\mathrm{GHz}$ detuning from the $v' = 0$ state.  At $3.75~\mathrm{mW}$, $\Omega_{\rm m}/\Omega_{\rm a}=0.016$, near the theory prediction of $0.013$.  The measured Rabi rate $\Omega_{\rm R}$ is only 63~\% of $\Omega_{\rm m} \Omega_{\rm a}/(2\Delta)$ due to interference from further-detuned Raman processes.
    \label{tab:f-det:fit}}
\end{table}

Figure~\ref{f-det}c shows the power dependence of $ \Omega_{\rm m} $ and $ \Omega_{\rm a} $, where $ \Omega_{\rm m} $ scales as $ P^{1/2} $ as expected. As discussed previously, the scaling of $ \Omega_{\rm a} $ is $P^{7/8}$ for weakly interacting particles.
However, due to the strong interaction between the two atoms in the $\ket{\uparrow_{\text{Na}}\downarrow_{\text{Cs}}}$ state, this approximation breaks down.
Coupled-channel calculations show that the wavefunction scaling
is well approximated by $P^{0.29}$ within the range of confinement in our experiment and the expected scaling of $ \Omega_{\rm a} \propto P^{0.79} $ agrees with the data.

While the measured single-photon Rabi frequencies of Table~I are in broad agreement with calculations (see Supplementary Materials), calculated scattering rates of the molecular and atomic state underestimate the decoherence in Fig.~\ref{f-raman}b.
From experimental measurements, the Raman transfer efficiency is limited by the molecular lifetime, together with a reduction in the Raman Rabi frequency due to destructive interference with intermediate states beyond $v'=0$ (see Table~~\ref{tab:f-det:fit}). We discuss loss and decoherence below.

The molecular lifetime measurement in Fig.~\ref{f-lifetime}a is performed
by preparing the molecule with a $\pi$-pulse,
followed after a variable delay by a second dissociating $\pi$-pulse. The measured lifetime of $0.196(14)~\mathrm{ms}$ is consistent with the decaying time scale of  $0.203(13)~\mathrm{ms}$ of the  Rabi oscillation
in Fig.~\ref{f-raman}b.
Preliminary experiments and theoretical considerations indicate that
the molecular lifetime may be limited by two-photon coupling
to the atomic continuum~\cite{YichaoYu}.
Atom loss is shown to be small in Fig.~\ref{f-lifetime}b
by measuring the two-body lifetimes of the atoms directly
without the second Raman frequency.
In principle, destructive interference that reduces the Raman Rabi rate $\Omega_{\rm R}$
for negative detunings $\Delta$ changes to constructive interference
for positive detunings, but additional molecular resonances make the positive region unusable.
More negative detunings that might reduce the scattering rate
were prevented by vanishing of $\Omega_{\rm R}$.

Separately we observe a decrease in coherence by a factor of 2 without laser spectrum filters,
suggesting that spectral impurity of the laser can be a significant source of loss. While we have not fully characterized the sources
of broadband noise, possibilities include amplified spontaneous emission (ASE) from the laser and fiber nonlinearities. Other potential decoherence sources include
fluctuations of the tweezer intensity and magnetic field, although the shape of the Rabi oscillation in Fig.~\ref{f-raman}b indicates that loss rather than frequency fluctuations are responsible for decoherence.
Based on the ratio between $\Omega_{\rm R}$ and the light shift,
the requirement on the tweezer intensity stability is $1\mathrm{\%}$ at $3.75~\mathrm{mW}$ power. We stabilize the power to $0.1\mathrm{\%}$, indicating that in the absence of beam-waist fluctuations, light shift is not a major source of decoherence.
Similarly, the measured Zeeman shift of $42.2(2)~\mathrm{kHz/G}$
does not cause significant decoherence for the measured magnetic field
fluctuation of $1.5~\mathrm{mG}$.

In conclusion, we have coherently formed a weakly bound NaCs molecule in an optical tweezer
by optical Raman transfer.  This process is enabled by utilizing a deeply-bound intermediate state, as well as highly-interacting initial atomic states.  A theoretical investigation including 8 excited molecular electronic potentials,
the electronically excited atomic continuum, and coupled-channel ground-state wavefunctions indicates
the potential for higher transfer efficiency than the observed value of $69$~\%.  Future experiments may benefit from better balancing of the up-leg and down-leg Rabi frequencies, for example by driving to more deeply bound states.  If possible, destructive interference that reduces the two-photon Rabi rate should be avoided.  Nonlinear optical effects that limit the molecular state lifetime can also be explored.

Our technique can be applied to form a diverse set of molecular species,
because it does not rely on a magnetic Feshbach resonance, states bound by only a few MHz,
or a narrow excited state. The formation of weakly bound molecules is a key step
in forming rovibrational ground-state molecules. By scaling up to many optical tweezers~\cite{Endres2016, Barredo2018,  PhysRevLett.122.203601}, large arrays with arbitrary geometry of highly controlled molecules can be achieved. These molecules comprise a flexible platform for quantum simulation,  quantum computing, precision measurements applications.

\begin{acknowledgments}
  We thank Bo Gao and Paul Julienne for discussion and Robert Moszynski for providing theoretical transition dipole moments of NaCs. This work is supported by the NSF~(PHY-1806595), the AFOSR~(FA9550-19-1-0089), ARO DURIP (W911NF1810194) and the Arnold and Mabel Beckman foundation. J.~T.~Z. is supported by a National Defense Science and Engineering Graduate Fellowship. W.~C. is supported by a Max Planck-Harvard Research Center for Quantum Optics fellowship. K.~W. is supported by an NSF GRFP fellowship. J.~M.~H. is supported by the U.K. Engineering and Physical Sciences Research Council (EPSRC) Grants No.\ EP/N007085/1, EP/P008275/1 and EP/P01058X/1. R.~G.~F. acknowledges financial support of the Spanish Project FIS2017-89349-P (MINECO), and the Andalusian research group FQM-207.
\end{acknowledgments}

\bibliography{master_ref}
\bibliographystyle{apsrev4-2}
\end{document}


\title{Coherent optical creation of a single molecule -- Supplemental material}
\author{Yichao~Yu}
\thanks{Y.Y.~and K.W.~contributed equally to this work.}
\email{yichaoyu@g.harvard.edu}
\gradstudent
\author{Kenneth~Wang}
\thanks{Y.Y.~and K.W.~contributed equally to this work.}
\gradstudent
\author{Jonathan~D.~Hood}
\affiliation{Department of Chemistry, Purdue University, West Lafayette, Indiana, 47906, USA}
\author{Lewis~R.~B.~Picard}
\gradstudent
\author{Jessie~T.~Zhang}
\gradstudent
\author{William~B.~Cairncross}
\harvardccb
\harvardphysics
\cua
\author{Jeremy~M.~Hutson}
\affiliation{Joint Quantum Centre Durham-Newcastle, Department of Chemistry, Durham University, Durham, DH1 3LE, United Kingdom}
\author{Rosario Gonzalez-Ferez}
\affiliation{Instituto Carlos I de F\'{\i}sica Te\'orica y Computacional, and Departamento de F\'{\i}sica At\'omica, Molecular y Nuclear,  Universidad de Granada, 18071 Granada, Spain}
\itamp
\author{Till Rosenband}
\affiliation{Agendile LLC, Cambridge, Massachusetts 02139, USA}
\author{Kang-Kuen~Ni}
\email{ni@chemistry.harvard.edu}
\harvardccb
\harvardphysics
\cua

\date{\today}

\maketitle

\section{3 Level Raman Transfer with Cross Coupling}

In our system, the separation in energy between the initial and target state is much smaller than the single photon detuning $\Delta$ from the intermediate state; see Fig.~1a in the main text.
Thus, there is significant cross coupling for the scattering rate and light shift, where each state is coupled to the excited state by the power in both beams.
Under these conditions, the scattering rate and light shift are proportional to the total power $P_{\text{tot}} = P_{\rm m} + P_{\rm a}$, where $P_{\rm m} (P_{\rm a})$ is the power in the beam that addresses the molecular (atomic) state.
The Raman Rabi frequency is proportional to $\sqrt{P_{\rm m}P_{\rm a}} $.
With a fixed total power and thus fixed scattering rate, the Raman Rabi frequency is maximized when $ P_{\rm m} = P_{\rm a} = P_{\text{tot}}/2$.

For the purposes of this work, we denote $\Omega_{\rm a} $($\Omega_{\rm m}$) as the Rabi frequencies to drive a transition from the atom (target molecule) state to the intermediate state used in the Raman transition corresponding to the power in a single beam, $P_{\rm a}$ ($P_{\rm m}$).
Thus, the Raman Rabi frequency is $\Omega_{\rm a}\Omega_{\rm m}/(2\Delta)$, where $ \Delta $ is the single photon detuning.
Cross coupling affects the scattering rate and light shift, and assuming equal powers in both beams, these quantities become $ \Gamma_{\rm e} \Omega_{\rm m}^2 / (2\Delta^2) + \Gamma_{\rm e} \Omega_{\rm a}^2 / (2\Delta^2) $ and $ \Omega_{\rm m}^2/(2\Delta) - \Omega_{\rm a}^2/(2\Delta) $, where $\Gamma_{\rm e}$ is the excited state linewidth. Since $ \Omega_{\rm m} >> \Omega_{\rm a} $, the scattering rate and light shift can be approximated as $ \Gamma_{\rm e} \Omega_{\rm m}^2/(2\Delta^2)$ and $\Omega_{\rm m}^2/(2\Delta)$.

\section{Raman Transfer with Many Intermediate States} \label{sm:sect_2}
With multiple intermediate states, the total Raman Rabi frequency, scattering rate, and light shift become a sum over all these possible states. In our model, we consider all vibrational states of $ \mathrm{c^3\Sigma^+(\Omega = 1)} $ and states in the atomic continuum as intermediate states. The Rabi frequency between two states is proportional to the electronic dipole moment, arising from the electronic wavefunctions, and the overlap of the vibrational wavefunctions. Under the assumption that the electronic dipole moment $ d $ is the same for all transitions between the ground state and the excited electronic state, the total Raman Rabi frequency becomes
\begin{equation}
  \Omega_{\rm R} = \frac{d^2E^2}{2\hbar^2} \displaystyle\sum_{v'} \frac{\braket{a|v'}\braket{v'|m}}{\Delta_{v'}},
\end{equation}
where $\ket{a}$ ($\ket{m}$) are the relative coordinate wavefunctions for the trapped atomic (target molecular) state, and $\ket{v'}$ is the relative-coordinate wavefunction for the excited state which includes both bound molecular states and the atomic continuum. $\Delta_{v'}$ is a single photon detuning with respect to the $\ket{v'}$ state, $ E $ is the electric field for the power in the single beam given a waist size. We can divide this sum into a sum over all the bound molecular states, and a sum over atomic continuum states,
\begin{equation}
  \Omega_{\rm R} = \frac{d^2E^2}{2\hbar^2}\displaystyle\sum_{v' \leq v'_{\text{max}}} \frac{\braket{a|v'}\braket{v'|m}}{\Delta_{v'}} +  \frac{d^2E^2}{2\hbar^2}\displaystyle\sum_{v' > v'_{\text{max}}} \frac{\braket{a|v'}\braket{v'|m}}{\Delta_{v'}},
\end{equation}
where $ v'_{\text{max}}$ corresponds to the excited molecular state, which is most weakly bound. The first term can be calculated numerically given an excited state potential. To calculate the second term, we first make an assumption that we are far enough detuned from the contributing continuum states that they are all at approximately the same detuning, $ \Delta_{\text{thresh}}$. This allows us to pull it out of the sum, and we obtain
\begin{equation}
  \Omega_{\rm R} \approx \frac{d^2E^2}{2\hbar^2}\displaystyle\sum_{v' \leq v'_{\text{max}}} \frac{\braket{a|v'}\braket{v'|m}}{\Delta_{v'}} +  \frac{d^2E^2}{2\hbar^2\Delta_{\text{thresh}}}\displaystyle\sum_{v' > v'_{\text{max}}} \braket{a|v'}\braket{v'|m} . \label{rabifreq}
\end{equation}
From requiring completeness and orthogonality,
\begin{equation}
  \displaystyle\sum_{v'} \braket{a|v'}\braket{v'|m} = \braket{a|m} = 0,
\end{equation}
we derive an expression for the second sum in equation \ref{rabifreq}
\begin{equation}
  \displaystyle\sum_{v'> v'_{\text{max}}} \braket{a|v'}\braket{v'|m} = -\displaystyle\sum_{v' \leq v'_{\text{max}}} \braket{a|v'}\braket{v'|m},
\end{equation}
which can be calculated from the vibrational states of the excited state potential. Thus, the Raman Rabi frequency can be approximated as:
\begin{equation}
  \Omega_{\rm R} \approx \frac{d^2E^2}{2\hbar^2}\displaystyle\sum_{v' \leq v'_{\text{max}}} \frac{\braket{a|v'}\braket{v'|m}}{\Delta_{v'}} -  \frac{d^2E^2}{2\hbar^2\Delta_{\text{thresh}}}\displaystyle\sum_{v' \leq v'_{\text{max}}} \braket{a|v'}\braket{v'|m}.
\end{equation}
We can account for the atomic continuum in the scattering rate in a similar way
\begin{align}
  \Gamma_{\rm s} &\approx \frac{d^2E^2}{2\hbar^2}\displaystyle\sum_{v' \leq v'_{\text{max}}}\Gamma_{v'} \frac{\braket{m|v'}\braket{v'|m}}{\Delta_{v'}^2} + \frac{d^2E^2\Gamma}{2\hbar^2\Delta_{\text{thresh}}^2}\displaystyle\sum_{v' > v'_{\text{max}}}\braket{m|v'}\braket{v'|m},
\end{align}
where $\Gamma_{v'}$ is the excited state linewidth for the $\ket{v'}$ state, and we have assumed the atomic continuum states all have the same linewidth $\Gamma$. Using normalization, we derive an expression for the sum in the second term
\begin{equation}
  \displaystyle\sum_{v' > v'_{\text{max}}}\braket{m|v'}\braket{v'|m} = 1 - \displaystyle\sum_{v' \leq v'_{\text{max}}}\braket{m|v'}\braket{v'|m}.
\end{equation}
We then arrive at an expression for the scattering rate, accounting for the atomic continuum,
\begin{equation}
  \Gamma_{\rm s} \approx \frac{d^2E^2}{2\hbar^2}\displaystyle\sum_{v' \leq v'_{\text{max}}}\Gamma_{v'} \frac{\braket{m|v'}\braket{v'|m}}{\Delta_{v'}^2} +  \frac{d^2E^2\Gamma}{2\hbar^2\Delta_{\text{thresh}}^2}\left(1 - \displaystyle\sum_{v' \leq v'_{\text{max}}}\braket{m|v'}\braket{v'|m}\right).
\end{equation}
We have replaced the full continuum with an effective state at the atomic threshold with strength proportional to the amount of wavefunction overlap between the target molecular state and all atomic continuum states. This model leads to a component in the scattering rate that is proportional to $ 1/\Delta_{\text{thresh}}^2$ and is smaller as one detunes farther from the threshold. We note that the first term can also scale nearly as $1/\Delta_{\text{thresh}}^2 $, if, as in our case, the largest terms in the summation are from states near the threshold.

A very similar result holds for the light shift $\delta$ and is written here for completeness

\begin{equation}
  \delta = \frac{d^2E^2}{2\hbar^2}\displaystyle\sum_{v' \leq v'_{\text{max}}} \frac{\braket{m|v'}\braket{v'|m}}{\Delta_{v'}} +  \frac{d^2E^2}{2\hbar^2\Delta_{\text{thresh}}}\left(1 - \displaystyle\sum_{v' \leq v'_{\text{max}}}\braket{m|v'}\braket{v'|m}\right).
\end{equation}

\section{Computational methods for bound and scattering states}

For scattering and near-threshold bound states, we solve the Schr\"odinger equation by
coupled-channel methods \cite{HUTSON1994}, using a basis set in which the electron spins
$s_{\rm Na}$ and $s_{\rm Cs}$ and the nuclear spins $i_{\rm Na}$ and $i_{\rm Cs}$ are separately
coupled,
\begin{equation}
  | (s_{\rm Na} s_{\rm Cs} S)
  (i_{\rm Na} i_{\rm Cs} I) F M_F\rangle.
  \label{eqbassif}
\end{equation}
Here $S$ and $I$ are the total electron and nuclear spins of the molecule or atomic pair. Their
resultant $F$ is the total spin, with projection $M_F$ onto the axis of the magnetic field. In
principle $F$ can be further coupled to the rotational angular momentum $L$, but in the present
work only functions with $L=0$ are included. In the absence of a magnetic field, $F$ and $M_F$ are
both good quantum numbers, but in a magnetic field only $M_F$ is conserved. The calculations in
this paper use basis sets with all possible values of $S$, $I$ and $F$ (for the required value of
$M_F$) that can be constructed with $s_{\rm Na} = s_{\rm Cs} = 1/2$,
$i_{\rm Na} = 3/2$ and $i_{\rm Cs} = 7/2$.

Scattering calculations are carried out using the MOLSCAT package \cite{molscat:2019,
  mbf-github:2020} and bound-state calculations using the related BOUND package
\cite{bound+field:2019, mbf-github:2020}. The calculations use singlet and triplet interaction
potentials for Na+Cs obtained by adjusting the potentials of Docenko \emph{et al.}\
\cite{Docenko2006} at short range, using the same procedure as for KCs \cite{Groebner2017},
to reproduce the experimental binding energy of the least-bound triplet state \cite{Hood2019}
and the position of the s-wave Feshbach resonance near 864~G \cite{Zhang2020}. The atomic
nuclear hyperfine coupling and Zeeman effects are fully included, using the Hamiltonian described
in \cite{Hutson2008}.

The bound-state calculations were initially used to predict the binding energies of states that lie
within about 1~GHz of each atomic threshold of interest. At zero magnetic field, the states fall
into groups from each hyperfine state of the atomic pair, labeled by $(f_{\rm Na},f_{\rm Cs})$. For
example, for the $(f_{\rm Na},f_{\rm Cs}) = (2,3)$ states that are of principal interest here, each
group comprises 5 zero-field levels with $F=1,2,3,4$ and 5; one such group is spread out over a
range of energies between 290 MHz and 770 MHz below the $(f_{\rm Na},f_{\rm Cs}) = (2,3)$
threshold. In a magnetic field, each zero-field level splits into $2F+1$ Zeeman sublevels labeled
by $M_F$. For each sublevel we can calculate the dependence of the energy on magnetic field, in
order to choose a target state for which the Raman transition frequency depends only weakly on
field.

The state that we ultimately chose as the target state for Raman transfer is the least-bound
$(F,M_F)=(5,5)$ sublevel with $(f_{\rm Na},f_{\rm Cs}) = (2,3)$, which was predicted at a binding
energy 763 MHz below the $(f_{\rm Na},f_{\rm Cs}) = (2,3)$ threshold. As described in the text, it
was subsequently located experimentally at a binding energy of 770.57150(9) MHz.

Multichannel bound-state wavefunctions, with components labeled by $S$, $I$ and $F$, are calculated
using the method of Thornley and Hutson \cite{Thornley1994}. For molecular bound states, they are
obtained on a grid of internuclear distances from 2.4~\AA\ to 50~\AA. For states of confined atomic
pairs, a harmonic confining potential is added to the molecular potentials and the grid is extended
to 3000~\AA.

\section{Excited Molecular States} \label{sm:excited_states}
To obtain a full calculation for the expected Raman Rabi frequency, light shift and scattering, we need to account for all the excited molecular states.
The lowest eight excited molecular potentials in NaCs asymptote to the dissociation limit of Na 3S + Cs 6P.
These eight potentials can be written in the Hund's case (a) basis as $ \mathrm{c^3\Sigma^+_{\Omega = 0^-},c^3\Sigma^+_{\Omega = 1}, b^3\Pi_{\Omega = 0^-}, b^3\Pi_{\Omega = 0^+}, b^3\Pi_{\Omega = 1}, b^3\Pi_{\Omega = 2}, B^1\Pi_{\Omega = 1}, A^1\Sigma^+_{\Omega = 0^+}}$ \cite{Korek2007}.
Due to spin-orbit coupling, especially large in Cs, the excited molecular states are mixed, containing components in multiple potentials.
Fortunately, single-channel potential energy curves for $\mathrm{c^3\Sigma^+_1}$ and $  B^1\Pi_1 $ have been experimentally determined, and are predictive across the entire energy range of the potential \cite{Grochola2010, Grochola2011, Liu2017}.
Since there is only one $ \Omega = 2$ potential at this dissociation limit, a single-channel potential has been determined for $ b^3\Pi_2$ as well \cite{Zabawa2012}.
Unfortunately, single-channel potentials for the other electronic states do not exist.
A deperturbation analysis was required for states belonging to the mixed $\mathrm{A-b}$ complex,
and the fitted potentials and spin-orbit coupling coefficients are only accurate within the energy range of the measured experimental data \cite{Zaharova2009}.
Since our experiment attempts to work in a regime where the predominant Raman coupling comes from the $v' = 0$ state of $ \mathrm{c^3\Sigma^+_1} $,
the contribution of the other seven potentials is mainly to a background that is nearly detuning independent.
Thus, for computational ease, we use the diabatic potentials for $ \mathrm{b^3\Pi_{0}, b^3\Pi_{1}} $ and $ \mathrm{A^1\Sigma^+_{0^+}}$.
Lastly, to our knowledge, no potential fit to experimental data exists for the $ c^3\Sigma^+_{0^-}$ state. Assuming that its contribution is only to the background, as an approximation, we simply double the contribution of all states except $v' = 0$ from $ c^3\Sigma^+_1$.

\section{Calculating Matrix Elements with Coupled Channel Ground State}
To calculate Rabi frequencies for transitions to the eight excited molecular potentials, we
obtain their wavefunctions from a single-channel calculation \cite{Fattal1996} using the potential energy curves described in section \ref{sm:excited_states}. We
then calculate overlap integrals between these functions and either the triplet ($S=1$) or singlet ($S=0$) components of the
coupled-channel wavefunctions for either the target molecular state or the confined atomic state.
These calculations assume that the intermediate states have pure singlet or triplet character and neglect its
nuclear spin character, which is unknown. These calculations include theoretical electronic transition dipole moments~\footnote{Private communication with R. Moszynski}.

\begin{figure}[ht!]
  \includegraphics[width=\textwidth]{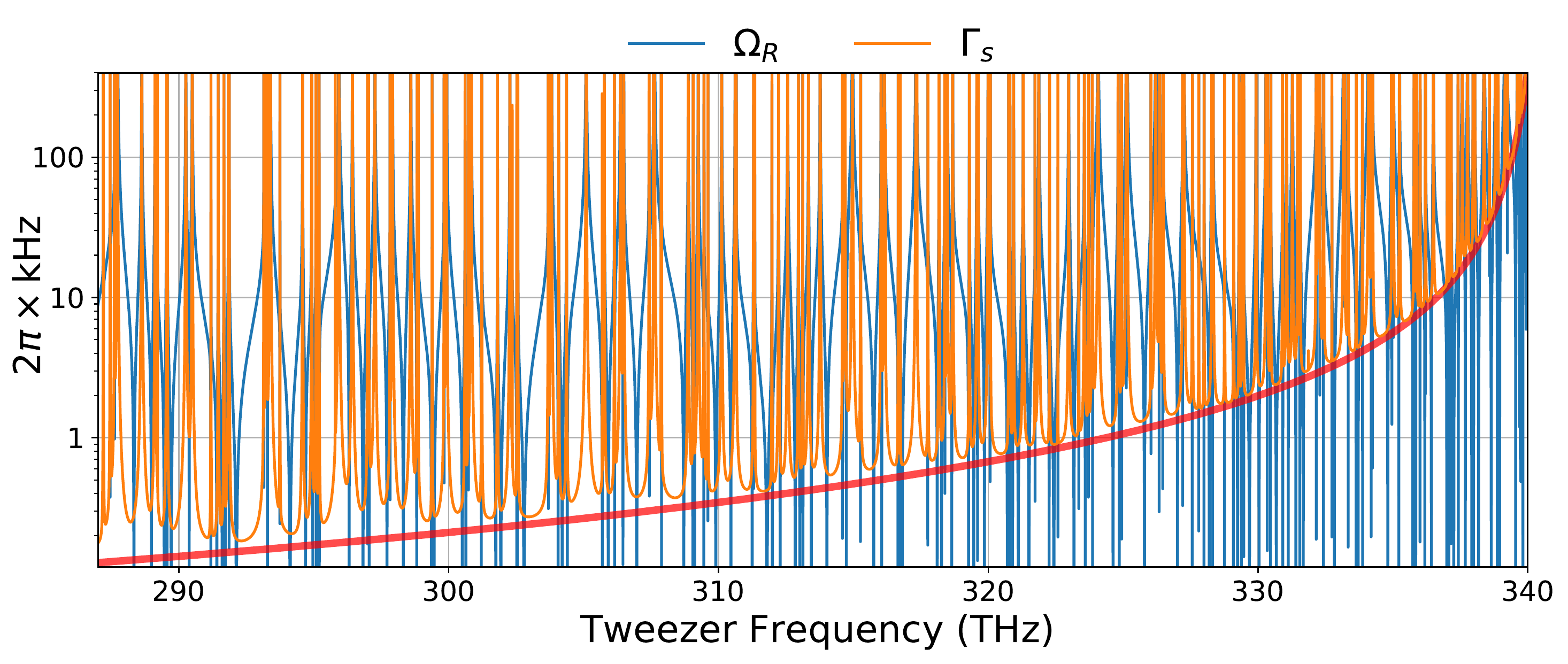}
  \caption{Full calculation of the Raman Rabi frequency and scattering rate
    as a function of the single-photon frequency
    including all vibrational states of $ \mathrm{c^3\Sigma^+(\Omega = 1)}$
    and the atomic continuum.
    The second frequency in the Raman transition is offset by about $770~\mathrm{MHz}$
    and resonant with the two photon transition.
    The assumed excited-state linewidth for all molecular lines is $50~\text{MHz}$.
    The red line shows the contribution that comes only from the near-threshold states.
    \label{f-sm}}
\end{figure}

\begin{figure}[ht!]
  \includegraphics[width=\textwidth]{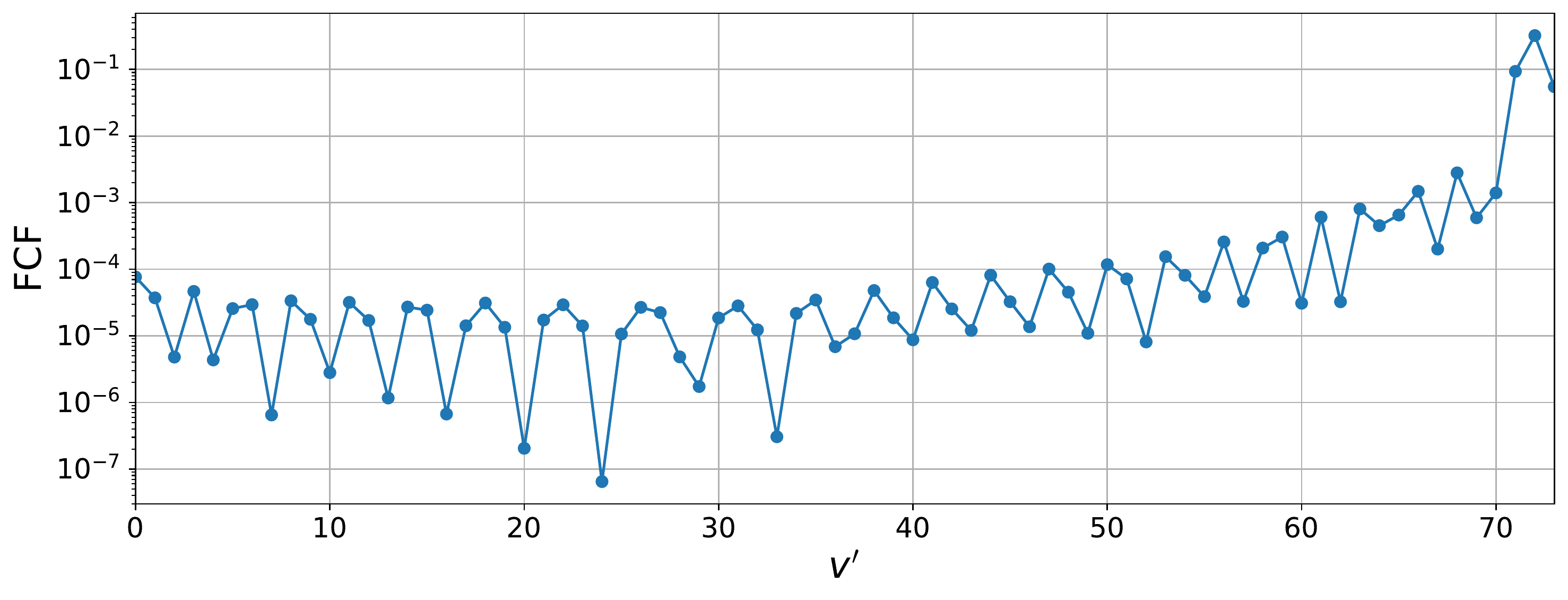}
  \caption{The Franck-Condon Factor (FCF) between the target molecular state and vibrational state $v'$ of $c^3\Sigma^+_1$. The potential supports 74 bound states, and the predominant contribution arises from the near-threshold bound states. The total overlap with all the bound states is $0.4822$, which is $96.8\%$ of the total triplet fraction of the target molecular state.
    \label{f-smfcf}}
\end{figure}

The full calculation of the Raman Rabi frequency and scattering rate using the methods in section \ref{sm:sect_2} is shown in figure \ref{f-sm}. The strong contribution from the atomic threshold arises from the large overlap to the near-threshold bound states shown in figure \ref{f-smfcf} for the $c^3\Sigma^+_1$ state. This leads to a $1/\Delta_{\text{thresh}}^2$ dependence of the scattering rate and favors using deeply bound states as the intermediate state for the Raman transfer. A similar result holds for the other excited state potentials.

\section{Effective Matrix Elements to a Single State}

In this calculation, there is a matrix element, $\Omega_{\mathrm{a}ij}, \Omega_{\mathrm{m}ij}$ between each channel $ i $ of the coupled channel ground state and an excited state $ j $. The total Raman Rabi frequency $ \Omega_{\rm R}$, light shift $ \delta$ and off-resonant scattering $ \Gamma_{\rm s} $ contribution from this single excited state is calculated as

\begin{align}
  \Omega_{\mathrm{R}j} &= \displaystyle\sum_{i} \frac{\Omega_{\mathrm{a}ij} \Omega_{\mathrm{m}ij}}{2\Delta}, \\
  \delta_j &= \displaystyle\sum_{i} \frac{\Omega_{\mathrm{a}ij}^2 - \Omega_{\mathrm{m}ij}^2}{2\Delta}, \\
  \Gamma_{\mathrm{s} j} &= \displaystyle\sum_{i} \Gamma_{\rm e} \frac{\Omega_{\mathrm{a}ij}^2 + \Omega_{\mathrm{m}ij}^2}{2\Delta^2},
\end{align}
where $ \Delta $ is the single photon detuning and $ \Gamma_{\rm e} $ is the excited state linewidth. The experiment, through measuring the detuning dependence of the Raman Rabi frequency and the resonance frequency, extracts a single $ \Omega_{\rm m} $ and $ \Omega_{\rm a} $. In order to compare to the theory, we introduce effective $ \Omega_{\rm a}', \Omega_{\rm m}', \Gamma_{\rm e}'$, so that $ \Omega_{\rm R} = \Omega_{\rm a}'\Omega_{\rm m}'/(2\Delta) $, $\delta = (\Omega_{\rm a}'^2 - \Omega_{\rm m}'^2)/(2\Delta) $, and $\Gamma_{\rm s} = \Gamma_{\rm e}'(\Omega_{\rm a}'^2 + \Omega_{\rm m}'^2)/(2\Delta^2) $. The effective Rabi frequencies for the $v' = 0$ state of $ c^3\Sigma^+_1$ and their ratio are reported in Table~\ref{tab:sm} at $3.75~\mathrm{mW}$ of tweezer power.


\begin{table}[ht]
  \centering
  \begin{tabular}{|c|c|} \hline
    $\Omega_{\rm m}'$ &  $2\pi \times 573 ~\mathrm{MHz}$ \\
    $\Omega_{\rm a}'$ &  $2\pi \times 7.37 ~\mathrm{MHz}$ \\
    $\Omega_{\rm a}'/\Omega_{\rm m}'$ & $ 0.013 $ \\ \hline
  \end{tabular}
  \caption{ The effective Rabi frequencies for the $v' = 0$ state of $c^3\Sigma^+_1$ at $ 3.75~\mathrm{mW}$ of tweezer power.
    \label{tab:sm}}
\end{table}

\section{Scaling of atomic matrix element with tweezer power}

Since our coupled-channel calculations include the harmonic confinement, we can calculate the enhancement in $\Omega_{\rm a}$ with increasing confinement. Using two calculations with $ \omega_{\text{trap}} = 36.5~\mathrm{kHz}$ and $\omega_{\text{trap}} = 80~\mathrm{kHz}$ trapping frequencies, and assuming a power law scaling, we find that the wavefunction overlap scales as $ \omega_{\text{trap}}^{0.58} = P^{0.29} $. Thus, $\Omega_{\rm a} \propto P^{0.79} $, which was also experimentally verified. However, we note that in the actual experiment, our system does not have spherically symmetric confinement. The radial trapping frequency is about $5.7$ times larger than the axial trapping frequency.
\todo{
  Power/intensity calibration
}

\bibliography{master_ref}